%% file: bbH_as_HH_background.tex
\documentclass[12pt,a4paper]{article}
\pdfoutput=1
\usepackage[utf8]{inputenc}
\usepackage{setspace}
\usepackage[table,dvipsnames]{xcolor}
\usepackage{epsf,amsmath,amssymb,graphicx,dcolumn}
\usepackage{caption}
\usepackage{subfig}
\usepackage{scalefnt,ulem,pstricks}
\usepackage{booktabs,multirow,tabularx}
\usepackage[colorlinks=true,allcolors={blue!70!black},linktocpage]{hyperref}
\usepackage{cleveref}
\usepackage{rotating}
\usepackage{microtype}
\usepackage{makecell}
\usepackage{cancel}
\usepackage{afterpage}
\usepackage[titletoc,title]{appendix}
\usepackage[numbers,sort&compress]{natbib}
\usepackage{amsmath,amsfonts,amsthm,bm}
\makeatletter \renewcommand{\@dotsep}{10000} \makeatother
\input{definitions.tex}

\begin{document}
\begin{titlepage}
\begin{flushright}
\vspace*{-1.5cm}
MPP-2023-103, PSI-PR-23-26, TIF-UNIMI-2023-16
\end{flushright}
\vspace{0.cm}

\begin{center}
{\Large \bf Taming a leading theoretical uncertainty in \boldmath{$\rm HH$} \\[0.3cm] \mbox{measurements via accurate simulations for \boldmath{b$\bar{\rm b}$H} production}}
\end{center}

\begin{center}
  {\bf Stefano Manzoni$^{(a)}$}, {\bf Elena Mazzeo$^{(b)}$}, {\bf Javier Mazzitelli$^{(c)}$},\\[1ex]
  {\bf Marius Wiesemann$^{(d)}$}, and {\bf Marco Zaro$^{(b)}$}\\[2ex]

$^{(a)}$ CERN, CH-1211 Geneva 23, Switzerland\\
$^{(b)}$ Universit\`a degli Studi di Milano \& INFN, Sezione di Milano, Via Celoria 16, 20133 Milano, Italy\\
$^{(c)}$ Paul Scherrer Institut, 5232 Villigen PSI, Switzerland\\
$^{(d)}$ Max-Planck-Institut f\"ur Physik, F\"ohringer Ring 6, 80805 M\"unchen, Germany\\[1ex]

\href{mailto:stefano.manzoni@cern.ch}{\tt stefano.manzoni@cern.ch},
\href{mailto:elena.mazzeo@cern.ch}{\tt elena.mazzeo@cern.ch},
\href{mailto:javier.mazzitelli@psi.ch}{\tt javier.mazzitelli@psi.ch},
\href{mailto:marius.wiesemann@mpp.mpg.de}{\tt marius.wiesemann@mpp.mpg.de},
\href{mailto:marco.zaro@mi.infn.it}{\tt marco.zaro@mi.infn.it}

\end{center}

\vspace{0.1cm}
\begin{center} {\bf Abstract} \end{center}\vspace{-0.5cm}
\begin{quote}
\pretolerance 10000

We present a new simulation for Higgs boson production in association with bottom quarks (\bbH{}) at next-to-leading order (NLO) accuracy matched to parton showers
in hadronic collisions.
Both contributions, the standard one proportional to the bottom-quark Yukawa coupling and the loop-induced one proportional
to the top-quark Yukawa coupling from the gluon-fusion process, are taken into account in a scheme with massive bottom quarks. 
Therefore, we provide the full simulation of the \bbH{} final state in the Standard Model, 
which constitutes also a crucial background to measurements for Higgs-boson pair ($HH$) production at the Large Hadron Collider when at least one of the 
Higgs bosons decays to bottom quarks. So far, the modeling of the \bbH{} final state induced one of the dominant theoretical uncertainties to $HH$ measurements, 
as the gluon-fusion component was described only at the leading order (LO) with  uncertainties of $\mathcal{O}(100\%)$. Including NLO corrections in its 
simulation allows us to reduce the scale dependence to $\mathcal{O}(50\%)$ so that it becomes subdominant with respect to other systematic uncertainties.
As a case study, we provide an in-depth analysis of the \bbH{} background to $HH$ measurements with realistic selection cuts in the $2b2\gamma$ channel. We also
compare our novel simulation with the currently-employed ones, discussing possible issues and shortcomings of a scheme with massless bottom quarks. Finally, we propagate the effect of the new \bbH{} simulation to $HH$ searches in the $2b2\gamma$ and $2b2\tau$ final states, and we find an improvement of up to 10\% (20\%) on the current (HL-LHC) limits on the $HH$ cross section.

\end{quote}

\end{titlepage}

\hrule
\renewcommand{\contentsname}{\normalsize Contents}
\tableofcontents{}

\vspace*{0.6cm}
\hrule
\vspace*{0.6cm}

\section{Introduction}
\label{sec:intro}

The measurement of the properties of the Higgs boson ($H$), discovered a decade ago~\cite{ATLAS:2012yve,CMS:2012qbp} at the Large Hadron Collider (LHC), is one of the major quests of the particle physics community.
Being naturally the least explored sector of the Standard Model (SM), its characterization is of utmost importance in the search for new-physics phenomena.
Current measurements of its coupling to top ($t$) and bottom ($b$) quarks, $W$ and $Z$ bosons, and tau leptons show a picture fully consistent with the SM expectation \cite{ATLAS:2022vkf,CMS:2022dwd}.
With the continuous increase of data taken at the LHC, those measurements will progressively become more precise, improving their sensitivity to deviations with respect to the SM.
On the other hand, further couplings will become accessible, which are currently restricted due to large statistical uncertainties.
First and foremost, this will be the case for the self interaction of the Higgs boson.

Through the mechanism of spontaneous electroweak symmetry breaking, the Higgs field generates not only the masses of the other particles in the SM, but also its own mass.
Crucial to this mechanism is the shape of the Higgs boson potential, which after symmetry breaking induces a self interaction for the physical Higgs boson.
In the SM the strength of this self interaction, $\lambda_{HHH}$, is fully determined by the Higgs mass and its vacuum expectation value.
An experimental determination of this coupling will therefore provide an essential verification %validation
whether this fundamental prediction of the SM is realized in nature, and it will be our first exploration of the Higgs boson potential.
Indeed, several new-physics models suggest a modification of the Higgs boson potential and the strength of the self interaction, so that any deviation from the 
SM expectation would be a clear signal of new physics.

The main avenue for the extraction of the Higgs self interaction is the measurement of Higgs boson pair  ($HH$)  production  (for a review, see \citere{DiMicco:2019ngk}).
At the LHC, the main production mechanism is through gluon fusion, similarly to the case of single-Higgs production.
Its measurement is extremely challenging, though, due to the very small production cross section of two Higgs bosons, which is three orders of magnitude smaller than for a single Higgs boson.
For example, the most recent measurement from the ATLAS experiment~\cite{ATLAS:2022jtk} limits the production cross section to be below 2.4 times the SM prediction ($\sigma^{HH}_{\rm SM}$) in the combination 
of all $HH$ search channels, and the self-coupling (by performing an 
exclusive $\lambda_{HHH}$ variation) is bound to the range $-0.6<\lambda_{HHH}<6.6$, both at $95\%$ confidence level.
With the increased volume of data at the end of the High-Luminosity phase of the LHC (HL-LHC), the Higgs-pair production cross section is expected to be measured with a significance of $4.9\sigma$ ($3.4\sigma$)~\cite{ATL-PHYS-PUB-2022-053}, when combining all search channels. The expected limits on the cross section will be reduced to $0.39(0.55) \times \sigma^{HH}_{\rm SM}$, and the determination of the Higgs self coupling is expected to be within $0.3(0.0)<\lambda_{HHH}<1.9(2.5)$, again at $95\%$ level, where the numbers shown in and outside the brackets include different 
assumptions on the systematic uncertainties~\cite{ATL-PHYS-PUB-2022-053}. Analogous results have been published by 
CMS~\cite{CMS:2022dwd,CMS:2022cpr,CMS:2022nmn,CMS:2022hgz,CMS:2022kdx}. 

Accurate theoretical predictions are crucial for the prospects of measuring Higgs boson pair production, and, more importantly, for the subsequent 
extraction of limits on $\sigma^{HH}_{\rm SM}$ as well as $\lambda_{HHH}$.
An impressive amount of effort has been devoted to the development of precise theoretical tools to predict the $HH$ cross section (see \citeres{deFlorian:2013uza,deFlorian:2013jea,Shao:2013bz,Frederix:2014hta,Maltoni:2014eza,Grigo:2014jma,deFlorian:2015moa,Borowka:2016ehy,Baglio:2018lrj,Grazzini:2018bsd,DeFlorian:2018eng,Bonciani:2018omm,Davies:2018qvx,Chen:2019fhs,Davies:2019djw,Baglio:2020wgt,Ajjath:2022kpv,Davies:2022ram,Muhlleitner:2022ijf} and references therein), 
notably complicated by its loop-induced nature.
A precise description of the signal, however, is not enough, and a good theoretical control over the relevant backgrounds of $HH$ measurements is equally 
important, as it is necessary not to lose significance in the signal extraction.

\begin{figure}[t]
  \centering
  \subfloat[$\yb{}$ diagram]
  {\includegraphics[height=.17\textwidth]{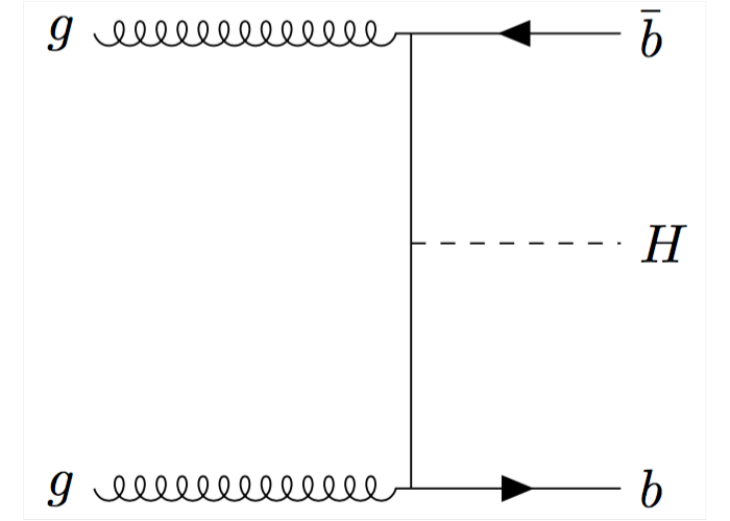}\label{fig:hbblo}}
  \hspace{3cm}
  \subfloat[$\yt{}$ diagram]
  {\includegraphics[height=.17\textwidth]{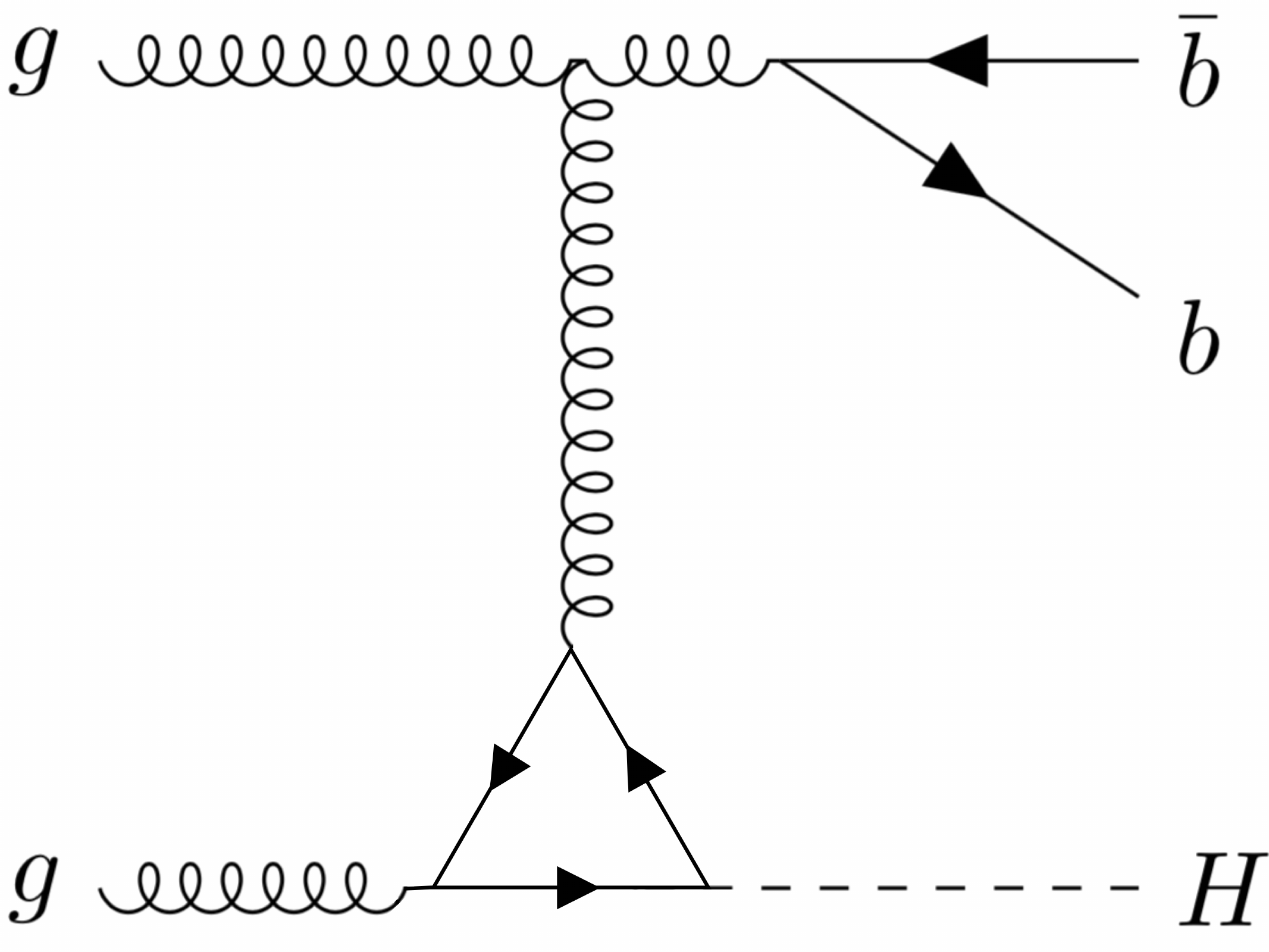}\label{fig:hbbnlot}}
\vspace{0.5cm}
  \caption{\label{fig:bbh}Sample Feynman diagrams for $b \bar b H$ production  
   proportional to \yb{} at tree-level and proportional to \yt{} induced by a top-quark loop.}
  \label{fig:hbblonlo}
\end{figure}

Given the very small $HH$ production rate, all of the phenomenologically relevant decay modes have (at least) one Higgs boson decaying 
into a pair of bottom quarks, since it has the highest branching ratio for a SM-like Higgs boson.
As a result, Higgs boson production in association with a bottom quark--antiquark pair ($b\bar{b}H$) is one of the irreducible backgrounds to all of these $HH$ searches at the LHC.

The $b\bar{b}H$ process receives contributions proportional to the bottom Yukawa coupling ($y_b$) where the Higgs couples to a bottom-quark line, see \fig{fig:bbh}\,(a), as well as contributions proportional to the top Yukawa coupling ($y_t$) where the Higgs boson couples 
to a closed top-quark loop, see \fig{fig:bbh}\,(b). The latter corresponds to the gluon-fusion process with an 
additional $b\bar{b}$ pair generated through QCD radiation.
Both mechanisms have comparably large production rates (with the $\ytsq{}$ contribution actually being dominant, see \citere{Deutschmann:2018avk}). The QCD perturbative corrections  are particularly large for both the $\ybsq{}$ contributions and 
the $\ytsq{}$ contributions, so that leading-order (LO) results do not provide an adequate description. We would like to point out that the $b\bar{b}H$ final state can be produced via other
production modes ($VH$ associated production with $V\to b\bar b$ and $b-$associated vector-boson fusion), with an impact on the cross section at the level of some/several tenths of percents~\cite{Pagani:2020rsg}. However, accurate
simulations for these production channels exist since long ago, and the inclusion of their contribution is thus trivial.

Since the bottom mass is neither particularly small nor particularly large compared to the typical scale of the $b\bar{b}H$ process,
suitable predictions can be obtained either in a four-flavour scheme (4FS) with massive bottom quarks or in a five-flavour scheme (5FS) 
where the bottom quark is treated as being massless.
In the 5FS, where the calculations are technically much simpler, significant progress has been made 
over the past years for the $\ybsq{}$ contribution \cite{Dicus:1998hs,Balazs:1998sb,Harlander:2003ai,Campbell:2002zm,Harlander:2010cz,Ozeren:2010qp,Harlander:2011fx,Buehler:2012cu,Belyaev:2005bs,Harlander:2014hya,Ahmed:2014pka,Gehrmann:2014vha,Duhr:2019kwi,Mondini:2021nck,Wiesemann:2014ioa,Krauss:2016orf,Ajjath:2019ixh,Ajjath:2019neu,Forte:2019hjc,Ahmed:2021hrf,Badger:2021ega}  with the cross sections at the third order in the strong coupling being the most remarkable advancement.
As far as 4FS calculations are concerned, there has been significantly less
progress in higher-order calculations for the $\ybsq{}$ contribution \cite{Dittmaier:2003ej,Dawson:2003kb,Dawson:2005vi,Liu:2012qu,Dittmaier:2014sva,Zhang:2017mdz,Wiesemann:2014ioa,Jager:2015hka,Krauss:2016orf,Pagani:2020rsg}, given the much more involved structure of the LO process,
with next-to-LO (NLO) QCD corrections (matched to parton showers) and combined with NLO electroweak ones, still being the state-of-the-art.
By contrast, a consistent study of both $b\bar{b}H$ production modes ($\ybsq{}$ and $\ytsq{}$) has been presented only in the 4FS by including NLO QCD 
corrections in all relevant coupling structures ($\ybsq{}$, $\ytsq{}$ and $y_b\,y_t$ interference contributions) \cite{Deutschmann:2018avk}, 
and it was shown that the NLO corrections to the $\ytsq{}$ contribution are similarly large as the LO $\ytsq{}$ (and $\ybsq{}$) cross section.
A substantial effort was also made in understanding differences between 4FS and 5FS cross sections, see e.g.\ \citeres{Maltoni:2012pa,Lim:2016wjo},
as well as  in obtaining consistent combinations of the two schemes \cite{Forte:2015hba,Forte:2016sja,Bonvini:2015pxa,Bonvini:2016fgf,Duhr:2020kzd}.
The advantages of either scheme in the context of simulating $b\bar{b}H$ production have been discussed in detail in \citere{Wiesemann:2014ioa}.
Throughout this paper we employ  the 4FS, owing to its better description of
differential observables related to final-state bottom quarks and
the definition of bottom-flavoured jets.

In the present paper, we build upon the NLO calculation of \citere{Deutschmann:2018avk} and match both $\ybsq{}$ and $\ytsq{}$ contributions
to parton showers (PS)  using the {\sc MadGraph5\_aMC@NLO} framework. We perform a detailed study at NLO, including PS-matching uncertainties, of the $b\bar{b}H$ background to Higgs boson 
pair production and show that including NLO corrections to the $\ytsq{}$ cross section allows us to tame the corresponding theoretical uncertainty in $HH$ 
measurements. To this end, we focus on the $2b2\gamma$ final state, and consider realistic selection cuts that are typically used in 
corresponding $HH$ analyses. We compare our results to the size of the $HH$ signal, as well as to the LO approximation of the $\ytsq{}$ terms based on the NNLOPS generator for inclusive Higgs boson production \cite{Hamilton:2012rf,Hamilton:2015nsa}, which is currently
employed by the experimental analyses. Finally, we propagate our improved $b\bar{b}H$ modeling to $HH$ searches performed by the ATLAS collaboration in the $2b2\gamma$ and $2b2\tau$ final states~\cite{ATLAS:2021ifb,ATLAS:2022xzm}, and we find an improvement of up to 10\% (20\%) on the current (HL-LHC) limits on $\sigma^{HH}_{\rm SM}$.

The paper is organized as follows: we describe our calculation in \sct{sec:outline}.
In \sct{sec:results} phenomenological  results for \bbH{} production are discussed
 in the context of $HH$ measurements. In particular, we start by describing the considered  
 setup (\sct{sec:setup}), then 
we study predictions for integrated and fiducial cross sections 
(\sct{sec:rates}), we analyze differential distributions (\sct{sec:distributions}), 
 we compare our new \bbH{} NLO+PS simulation in the 4FS with the NNLOPS prediction  and describe how to consistently combine them 
 (\sct{sec:comparison-with-NNLOPS}), and finally we study the impact of the new \bbH{} modeling on experimental $HH$ searches (\sct{sec:impact-on-hh-searches}). 
 The paper is summarized in \sct{sec:conclusions}.

\section{Outline of the calculation}
\label{sec:outline}

The calculation of the full NLO QCD corrections to $b\bar{b}H$ production has been discussed in detail in \citere{Deutschmann:2018avk}, which we refer to at this point.
Here, we briefly recall only the most relevant aspects.

The LO contribution to $b\bar{b}H$ production in the 4FS starts 
at $\calo{\alpha_s^2}$ in QCD perturbation theory, and is mediated by the
bottom-quark Yukawa coupling. Hence, the coupling structure of the LO process is
$\ybsq\,\alpha_s^2$. A sample Feynman diagram is shown in \fig{fig:bbh}\,(a).
The squared gluon-fusion diagram in \fig{fig:bbh}\,(b), where
the Higgs boson couples to a closed top-quark loop and which thus yields 
a contribution proportional to \ytsq{}, enters only at $\calo{\alpha_s^4}$, i.e.\ it is formally
only a next-to-NLO (NNLO) QCD correction to the LO \ybsq{} contribution. However, since this subprocess is enhanced 
by $\ytsq/\ybsq$ with respect to the LO process proportional to \ybsq, it involves a (potentially enhanced) $g\to b\bar b$ splitting,
and it receives large NLO QCD corrections (considerably larger than the \ybsq{} one \cite{Deutschmann:2018avk}),
the \ytsq{} contribution actually yields the dominant share to the \bbH{} cross section in the SM.
Interference contributions between the two processes exist in the 4FS, and they are formally part of the NLO QCD corrections to the 
$\ybsq$ induced LO process and thus enter at $\calo{\ybyt\,\alpha_s^3}$. However, when compared to the $\ybsq$ and $\ytsq$ contributions,
the size of the interference is much smaller, with effects at the few-to-ten percent level.

Due to their different coupling structures, \ybsq{}, \ybyt{}, and \ytsq{} terms define 
three separately gauge-invariant contributions under higher-order QCD corrections. 
As a result, we can compute the NLO QCD corrections for each of the three contributions 
individually and decompose the \bbH{} cross section up to $\calo{\alpha_s^5}$, i.e.\ formally 
next-to-NNLO (N$^3$LO) w.r.t.\ the LO process, as follows:
\begin{equation}
\begin{split}
    {\rm d} \sigma &    = \ybsq{}\,\alpha_s^2\left( \Delta_{\ybsq}^{(0)} + \alpha_s \Delta_{\ybsq}^{(1)} \right) +y_t y_b\, \alpha_s^3\left( \Delta_{\ybyt}^{(0)} + \alpha_s \Delta_{\ybyt}^{(1)} \right) + \ytsq{} \,\alpha_s^4\left( \Delta_{\ytsq}^{(0)} + \alpha_s \Delta_{\ytsq}^{(1)} \right)\, .
\end{split}
\label{eq:hbbxsec2}
\end{equation}

Since the NLO corrections to the contributions proportional to \yt{}
require the calculation of two-loop $2\to 3$ amplitudes with internal massive fermion
lines, which are beyond current technology, the approximation of a heavy top quark
is employed to derive those contributions at NLO. The effective field theory resulting 
from integrating out the top quark is well known, it has been employed for several 
processes and it has been implemented in a model~\cite{Artoisenet:2013puc,Demartin:2014fia} in the Universal Feynrules Output (UFO) format~\cite{Degrande:2011ua}, so that it can be employed through 
automated Monte Carlo tools, such as \textsc{MadGraph5\_aMC@NLO}~\cite{Alwall:2014hca, Frederix:2018nkq}.

The validity of the heavy top approximation for \bbH{} production has been studied and discussed
at length in \citere{Deutschmann:2018avk} for the \ytsq{} contribution. In fact, the case at hand 
corresponds essentially to Higgs+jet ($H$+$g$) production in gluon fusion with a
$g\to b\bar{b}$ splitting either in the initial or in the final state. For Higgs+jet 
production it has been shown in various publications, see 
for instance \citeres{Harlander:2012hf,Neumann:2014nha,Neumann:2018bsx,Jones:2018hbb,Chen:2021azt},
that the heavy top limit provides an excellent approximation to the exact cross section, especially for the computation of radiative corrections, and
even well beyond the naively expected validity range, as \citere{Chen:2021azt} has shown quite impressively.
Also for the \ytsq{} contribution to the \bbH{} process, \citere{Deutschmann:2018avk} comes to the conclusion
that at LO the heavy top-mass approximation works extremely well, with differences from the full computation 
below 10\% as long as the probed momentum scales (Higgs or leading b-jet transverse momentum, 
or invariant mass of the b-jet pair) do not exceed $200\,\GeV$.
Given the relatively small impact, we refrain from performing the Born-improved (BI) 
approximation of the \ytsq{} contribution here, which rescales the effective-field-theory result with the exact top-mass
dependence at LO, since the mass effects are expected to be far below the perturbative uncertainties in the 
signal region under consideration \cite{Deutschmann:2018avk}.

We have computed all NLO QCD corrections corresponding to the decomposition in \eqn{eq:hbbxsec2}
using the  \textsc{MadGraph5\_aMC@NLO} framework, including the matching to PS.
Events for the \ybsq{}, \ybyt, and \ytsq{} contributions can be either generated simultaneously or separately for each contribution, with the latter option resulting in a better numerical convergence.
Given that, as it has already been mentioned, the interference contribution is comparably small (about $-5$\% to $-10$\%),
and given the relatively large NLO 
scale uncertainties of the \ybsq{} and \ytsq{} contributions, we deem the $\ybyt$ contribution not 
relevant to the purpose of this paper and we refrain from including it here. 
We stress however that the $\ybyt$ interference can be computed/turned on within our implementation 
of the \bbH{} generator in \textsc{MadGraph5\_aMC@NLO}.

\section{Phenomenological results}
\label{sec:results}
\subsection{Setup}\label{sec:setup}
We have used the same settings as in \citere{Deutschmann:2018avk}, which allowed us
to validate our new NLO+PS implementation against the previous NLO results. In particular, we 
consider proton--proton collisions with 13\,TeV centre-of-mass energy and set following values
for the on-shell masses:
\begin{align}
    m_b = 4.92\, \GeV,\quad m_t = 172.5\, \GeV, \quad m_H = 125\,\GeV.
\end{align}
While also \yt{} is renormalized on-shell using the top-mass value above, for  
$y_b$ we use the $\overline{\rm MS}$ scheme, evaluated from an input value $m_b(m_b) =4.18\, \GeV$
and evolved to the respective central renormalization scale ($\mu_R^0$) with a four-loop running.
From that value $y_b(\mu_R^0)$ scale variations are determined with a two-loop evolution, consistent 
with the order of our calculation. This procedure follows the recommendation 
by the LHC Higgs cross section working group~\cite{deFlorian:2016spz}.
For the parton densities we use the NLO set of NNPDF3.1~\cite{NNPDF:2017mvq} with 4 flavours and the 
corresponding two-loop running of $\as$ (LHAID=320500).
The central factorization and renormalization scales are set to
\begin{equation}
    \mu_R^0 = \mu_F^0 = \frac{H_T}{4}= \frac{1}{4}\sum_{i\,\in\textrm{\,final}} \sqrt{m^2(i) + p_T^2(i)}\,,
\end{equation}
where the sum runs over all final state particles and $m(i)$, $p_T(i)$ are their respective mass, transverse momentum. The associated scale uncertainties are determined through the customary 9-point envelope obtained by varying independently the two scales up and down by a factor 2. 
The shower scale is chosen with a lower reference value $Q_{\rm sh}=\frac{H_T}{4}$ 
than the default one in \textsc{MadGraph5\_aMC@NLO},\footnote{The setup employed in this paper
    corresponds to setting {\tt shower\_scale\_factor=0.5} in the {\tt run\_card}.} as suggested in \citere{Wiesemann:2014ioa}.
Here, we also study shower-scale variations around the central value by a factor of two up and down, which are quoted
separately from those associated to the renormalization and factorization scales.

We have generated separately the contributions proportional to \ybsq{} and \ytsq{}, while, as pointed 
out before, we neglect the \ybyt{} interference, which has a subleading numerical impact, well within 
the scale uncertainties. For reference, we quote the total inclusive cross section in 
\tab{tab:ratesinc} separated by \ybsq{}, \ytsq{} and their sum.
As already observed in \citere{Deutschmann:2018avk}, the NLO QCD 
corrections in all cases are substantial, and they lead to a reduction of the scale uncertainties.
For the \ybsq{} contribution the correction amounts to more than $+50$\%, while for 
the \ytsq{} contributions we observe effects as large as $+140$\% at NLO QCD, 
rendering LO predictions completely unreliable.\footnote{We note that, while the same 
setup as \citere{Deutschmann:2018avk} was employed in our work, the quoted numbers are different, because 
of an update of the 4-flavour NNPDF3.1 set~\cite{NNPDF:2017mvq}, see also 
footnote 6 on page 11 in \citere{Deutschmann:2018avk}.}

\def\arraystretch{1.5}
\begin{table}[t]
    \centering
    \begin{tabular}{c|cc|cc}
        Contr. & LO & $\delta \mu_{R,F}$ & NLO & $\delta \mu_{R,F}$  \\
        \hline \hline
         \input{table_rates_inc}
    \end{tabular}
    \vspace{0.5cm}
    \caption{\label{tab:ratesinc} Inclusive cross section for $pp \to b \bar b H$. Numbers are in fb.}
\end{table}
\def\arraystretch{1.}

As a representative case of $HH$ searches, we consider here one of the
 most sensitive search channels,
where one Higgs boson decays into bottom quarks and the other decays into photons, assuming a $H\to \gamma\gamma$ branching 
ratio of ${\rm BR}(H\to\gamma\gamma)=0.227\%$~\cite{LHCHiggsCrossSectionWorkingGroup:2016ypw}.
As far as the fiducial setup is concerned, throughout this paper, we consider selection cuts motivated 
by a recent $HH$ search by the ATLAS collaboration~\cite{ATLAS:2021ifb}. In particular, we select anti-$k_T$~\cite{Cacciari:2008gp}
jets as implemented in {\sc FastJet}~\cite{Cacciari:2011ma} with $R=0.4$ and define bottom-flavoured jets
(short $b$-jets) as those containing at least one $B$ hadron with the requirements
\begin{equation}
    p_T(j) > 25\,\GeV\qquad {\rm and}\qquad |\eta(j)|<2.5\,,
\end{equation}
assuming a $b$-tagging efficiency of 100\% and without mistagging for our theoretical study. The $HH$
signal region is defined by requiring (exactly) two $b$-jets and two photons (the QED shower is disabled
in our simulations).
The invariant mass of the $b$-jet pair is required to be within 
\begin{equation}
    \label{eq:mbjcut}
    80\, \GeV < m(b_1,b_2)< 140\,\GeV\,.
    \footnote{We have applied this cut instead of the full discriminant employed by 
    the analysis of \citere{ATLAS:2021ifb}, since this 
invariant-mass region provides the highest score for the $HH$ signal.} 
\end{equation}
Notice that for the distribution in the number of $b$-jets this requirement is lifted.
The two photons must satisfy the following relations:
\begin{equation}
    105\,\GeV < m(\gamma_1, \gamma_2) < 160\,\GeV, \quad |\eta(\gamma_i)|< 2.37, \quad
    \frac{p_T(\gamma_1)}{m(\gamma_1, \gamma_2)} > 0.35, \quad \frac{p_T(\gamma_2)}{m(\gamma_1, \gamma_2)} > 0.25\,.
\end{equation}
In practice, since no detector effects are applied and no QED shower is included, we always have 
$m(\gamma_1, \gamma_2) - m_H = \mathcal{O}(\Gamma_H)$, so that the first requirement is trivially fulfilled.

Besides the above set of cuts, which we will refer to as \textit{fiducial cuts}, we define the variables
\begin{equation}
m_{2b2\gamma} = m(b_1,b_2,\gamma_1,\gamma_2)\,,
\end{equation}
and
\begin{equation}
\label{eq:mbbggs}
    \mbbggs = m_{2b2\gamma} - m(b_1,b_2) - m(\gamma_1,\gamma_2) + 2m_H \,,
\end{equation}
and we consider three possible categories for cuts on $\mbbggs$:
\begin{equation}
    \mbbggs < \infty, \qquad \mbbggs < 500\,\GeV, \qquad \mbbggs < 350\,\GeV \,.
\end{equation}
Thus, the first scenario corresponds to the fiducial cuts, and the others apply increasingly stronger requirements
on $\mbbggs$. These three regions provide complementary information on the 
Higgs potential. In particular, the region close to threshold has an enhanced sensitivity
to the trilinear Higgs coupling.

In the presentation of our phenomenological results we will compare to reference 
predictions for both the \ytsq{} \bbH{} background and the $HH$ signal.
The former is obtained from the NNLOPS generator for inclusive Higgs boson production
\cite{Hamilton:2012rf,Hamilton:2015nsa}, which is currently used to model 
the \ytsq{} \bbH{} background by the experiments, and it is formally LO+PS accurate
for that contribution. To this end, we have followed closely the corresponding 
simulation employed by ATLAS~\cite{ATLAS:2019nkf}. The NNLOPS generator merges $0$ and $1$-jet
multiplicities at NLO QCD using the MiNLO$^\prime$ \cite{Hamilton:2012rf} method and then reaches NNLO QCD
accuracy through reweighting to NNLO rapidity distribution. In addition, the sample
is normalized to the reference gluon-fusion Higgs production cross section~\cite{deFlorian:2016spz},
which includes the N$^3$LO corrections~\cite{Anastasiou:2016cez,Anastasiou:2014lda,Anastasiou:2014vaa,Anastasiou:2015vya}. The renormalisation and factorisation scales in the NNLOPS calculation are set to $\mu_R=\mu_F=m_H/2$, the PDF4LHC15~\cite{Butterworth:2015oua} parton densities are used, and {\sc Pythia8}~\cite{Bierlich:2022pfr} is employed
to perform the parton showering and to include the decay of the Higgs boson to two photons.

The $HH$ signal is simulated at NLO QCD including the full top mass effects~\cite{Heinrich:2017kxx}, using as reference value for the renormalisation and factorisation scales $\mu_R=\mu_F=m_{HH}/2$ and PDF4LHC15 as parton densities. The calculation is matched to {\sc Pythia8} to include  parton showering and to include the decay of the two Higgs bosons to bottom quarks and photons.
The $HH$ signal cross section is normalized to the value $\sigma^{\mathrm{SM}}_{\mathrm{ggF}} (pp \to HH) = 31.0^{+2.1}_{-7.2}$~fb from \citere{Grazzini:2018bsd}, which is obtained by combining the full NLO QCD calculation with the NNLO corrections obtained in the heavy-top limit and improved by including partial finite top mass effects via a suitable reweighting of the scattering amplitudes.

\subsection{Fiducial rates}
\label{sec:rates}
\afterpage{
\begin{table}[t]
    \centering
    \begin{tabular}{c||c|c|c|ccc||c|c}
        Cut & Contr. & Run & LO & NLO & $\delta \mu_{R,F}$ & $\delta Q_{sh}$ & \makecell{NNLOPS \\($y_t^2$ LO)} & \makecell{HH \\ signal}  \\
        \hline \hline
         \input{table_rates}
    \end{tabular}
    \vspace{0.5cm}
    \caption{\label{tab:rates} Rates for the process $pp \to b \bar b H$ with $H\to \gamma\gamma$ decay. Numbers are in ab.}
\end{table}
\clearpage
}

In this section, we study in detail the relevance of the accurate modeling of the \bbH{} production as a background to $HH$ searches. We start by  discussing total and fiducial rates for \bbH{} production including the decay $H\to \gamma\gamma$ 
in \tab{tab:rates}.
In particular, we show LO and NLO QCD predictions for the \ybsq{} and \ytsq{} contributions
as well as their sum for the fully inclusive cross section and in the three fiducial categories introduced in the 
previous section with different requirements on $\mbbggs$. All of these results are provided for both 
the matching to {\sc Pythia8}~\cite{Bierlich:2022pfr} and to {\sc Herwig7}~\cite{Bellm:2019zci}. In the case of {\sc Pythia8}, we also
present results obtained with the recently developed MC@NLO-$\Delta$ matching procedure~\cite{Frederix:2020trv}.
For our novel NLO+PS results we provide residual uncertainties due to variations of the factorization and renormalization
scales as well as separately due to variations of the shower scale.
For reference, we quote the cross section of the currently
adopted approximation of the \ytsq{} contribution $gg\to b\bar{b}H$ by the LHC experiments, which stems from 
the NNLOPS generator in \citeres{Hamilton:2012rf,Hamilton:2015nsa} and is 
effectively described only at LO and for massless bottom quarks, i.e.\ in the 5FS, in the heavy top approximation 
(reweighted by the exact $gg\to gH$ amplitude \cite{Hamilton:2015nsa}). In addition, we show the $HH$ signal
cross section in the different categories as a reference.

The main conclusions drawn from \tab{tab:rates} can be summarized as follows:
\begin{itemize}
\item The NLO QCD corrections are very large, and their inclusion becomes absolutely crucial in order to obtain
a reliable estimate of the \bbH{} background in any of the shown categories.
They are significantly larger for the \ytsq{} contributions, increasing the LO prediction by about $+150\%$, while the effect in the \ybsq{} ones is roughly $+50\%$.
The residual uncertainties from renormalization and factorization scale variations
are at the level of $+ 50\%$ and $- 30\%$ for the full \bbH{} cross section (sum of \ybsq{} 
and \ytsq{} terms) in all fiducial categories.
\item At NLO, the impact of using either {\sc Pythia8} (with or without the MC@NLO-$\Delta$ procedure) or {\sc Herwig7} as a PS, 
as well as uncertainties related to the variation of the shower scale, are
    at the level of 10\%. Thus, they can be considered as subleading
    with respect to the perturbative uncertainties. Only at LO, where the uncertainties of the predictions 
    are generally much more significant, the difference of using the {\sc Pythia8} or {\sc Herwig7} PS is more noticeable.
\item The fiducial cuts have a substantial impact on the size of the \bbH{} background.
With respect to the fully inclusive cross section, the baseline cuts reduce the cross section
by roughly a factor of 100. Further restrictions on $\mbbggs$ lead to a further 
decrease of the cross section by about $25\%$ ($55\%$) for $\mbbggs<500$\,GeV
($\mbbggs<350$\,GeV). It is interesting to notice that the inclusive \bbH{} cross section
is almost two orders of magnitude larger than the $HH$ signal, but the strong suppression  induced by the fiducial selection cuts on the \bbH{} background leads to
similarly large cross sections in the three cut scenarios, 
with the exception of the $\mbbggs<350$\,GeV category, where the $HH$ signal 
is more than a factor of two smaller than the \bbH{} background.
Regardless, the fact that the sum of \ybsq{} and \ytsq{} terms is at least as large 
as the $HH$ signal cross section in all categories, underlines the importance of 
an accurate modeling of the \bbH{} background at NLO QCD.
\item When we compare our results to the previously adopted NNLOPS simulation
for the \ytsq{} terms, which are effectively described only with LO accuracy in the 
matrix elements, we find large differences between our NLO+PS 4FS
calculation and the 5FS NNLOPS one, where the latter yields a cross section that
is about a factor of two larger. So far, in the ATLAS experiment analyses, the NNLOPS calculation
is assigned a $100\%$ uncertainty. Nevertheless, it is important to understand
the origin of these significant differences to judge which prediction can be trusted. 
We have traced back a substantial effect in the NNLOPS results to $g\to b\bar{b}$ splittings
in the PS, which make up almost half of the fiducial rates for the NNLOPS result,
and even more inclusively. The corresponding 
value when turning off $g\to b\bar{b}$ splittings in the PS is given below the nominal
NNLOPS prediction. We will discuss this point in more detail at the end of \sct{sec:distributions}, but we already
point out our main conclusions: when the PS generates hard radiation, in particular 
bottom quarks through $g\to b\bar{b}$ splittings, it acts outside its validity range and the 
results are subject to very large uncertainties. Even worse, we have checked that in the case at hand, the enhancement of the cross section due to the PS
originates almost entirely from events with exactly two bottom quarks coming 
from a gluon splitting, which thus creates two hard and separated $b$-tagged jets.
Not only are such topologies not 
accurately described by a PS in general (due to its underlying soft/collinear approximation), 
but also the same kinematical region should already be described by the LO $gg\to b\bar{b}H$ matrix element.
This suggests that by applying the PS to the inclusive $pp\to H$ NNLOPS 
sample, topologies with two hard $b$-jets may be double counted.
Note that, of course, the NNLOPS matching is consistent up to relative $\alpha_s^2$ accuracy (relative to the LO $gg\to H$ cross section), the same order where 
the perturbative series of the $\ytsq{}$-induced $\bbH$ cross section starts,
so that such a double counting is formally 
beyond the nominal accuracy of that prediction.
As a result, we conclude that the assigned $100\%$ uncertainty (and possibly more) applied to the NNLOPS 5FS prediction by 
the experiments so far is realistic,  due to its insufficiencies
in describing the bottom-quark/$b$-jet kinematics. By contrast, the 4FS calculation that 
describes these
contributions at the level of the hard matrix elements up to NLO QCD 
 provides a more accurate and realistic determination of the cross section with two 
bottom quarks/$b$-jets and their kinematics. Moreover, a clear reduction of the nominal 
uncertainties over the previously used NNLOPS can be achieved with the 4FS calculation.
\end{itemize}

Before turning to the description of differential distributions, we briefly comment on the amount of negative weights in our simulations. Negative 
weights are unavoidable in the MC@NLO matching~\cite{Frixione:2002ik} employed by {\sc MadGraph5\_aMC@NLO}. A large number of negative weights hampers
the quality and the performance of the simulations, in the sense that larger event samples have to be generated in order to attain the same 
statistical uncertainties of a simulation with solely positive weights. The need to reduce negative weights in {\sc MadGraph5\_aMC@NLO}
has motivated the development of new strategies, such as the MC@NLO-$\Delta$ approach \cite{Frederix:2020trv}. In \tab{tab:negw} we show the fractions of events with negative weights $F$ in our samples and make the following observations:
First, the fraction of negative weights has a mild dependence on the choice of shower-scale
in general. Second, for both the $\ybsq$ and $\ytsq$ contributions the following
patterns arise: $F_{\textrm{PY}8} > F_{\textrm{HW}7} > F_{\textrm{PY}8-\Delta}$.
Third, there is a noticeable 
improvement due to the usage of  MC@NLO-$\Delta$, which is particularly
relevant for the $\ybsq$ term, where one has $F_{\textrm{PY}8}^{\ybsq} - F^{\ybsq}_{\textrm{PY}8-\Delta}\simeq 5\%$. While there is a slightly smaller reduction for 
the $\ytsq$ contribution, $F$ is already smaller in that case using
 the standard MC@NLO matching. 
Although these numbers may look small, a reduction of 5\% in
the negative weight fraction can halve the computational cost of the simulation when $F\simeq 35\%$, as it is displayed in Fig.\,1 of \citere{Frederix:2020trv} for instance.

\begin{table}[t]
    \centering
    \begin{tabular}{c|c|ccc}
        Sample & $Q_{\textrm{sh}}$ & PY8 & PY8-$\Delta$ & HW7 \\
        \hline
        \hline
        \multirow{3}{*}{\ybsq{}} & $\times 2  $ & 39.1 & 33.8 & 36.8 \\
                                 & $\times 1  $ & 38.5 & 33.9 & 36.4 \\
                                 & $\times 0.5$ & 38.1 & 33.8 & 36.3 \\
        \hline
        \multirow{3}{*}{\ytsq{}} & $\times 2  $ & 33.0 & 28.8 & 30.7 \\
                                 & $\times 1  $ & 32.5 & 29.0 & 30.8 \\
                                 & $\times 0.5$ & 32.5 & 28.9 & 31.1 \\
    \end{tabular}
    \vspace{0.3cm}
    \caption{\label{tab:negw} Percentage of negative weights in the various samples generated.}
\end{table}

\subsection{Differential distributions}
\label{sec:distributions}
We now turn to studying differential distributions, which allow us to make further statements
on the relevant kinematics of the \bbH{} process. These results may be useful, for instance, 
to obtain a better understanding of which phase-space regions could be used in
 $HH$ measurements to reduce the \bbH{} background or, in general, what impact different
 cuts may have on the \bbH{} process. Moreover, we will analyze further the different 
 sources of uncertainties, including the differences observed using different PS or matching procedures.
 
 The figures throughout this section are organized as follows: there is a main frame 
 showing the absolute predictions as cross section per bin for the \ytsq{} and \ybsq{} contributions
 at NLO+PS using {\sc Pythia8} (orange, light blue) and using {\sc Herwig7} 
 (red, green). Then there are four ratio
 panels below. The first two show the envelope of the 9-point variations of renormalization and factorization
 scales for the \ytsq{} and \ybsq{} contributions at NLO+PS as a shaded band normalized
 to the central prediction, respectively. 
 In these two frames one can also appreciate the size NLO corrections, 
 as the respective LO+PS result is shown as a dashed curve.
 The last two ratio panels show, again for the \ytsq{} and \ybsq{} contributions respectively, 
shaded bands for the dependence on the shower scale for both {\sc Pythia8} 
and {\sc Herwig7}, using the same color scheme as the main inset. Besides these two set of curves,
predictions obtained with {\sc Pythia8} in conjunction with the MC@NLO-$\Delta$ matching are also shown (solid curves in brown and cyan), as well as the NLO fixed-order predictions  (dashed curves in magenta and dark blue). All results are normalized to the standard {\sc Pythia8} central prediction.

\begin{figure}[t]
    \centering
    \includegraphics[width=0.48\textwidth]{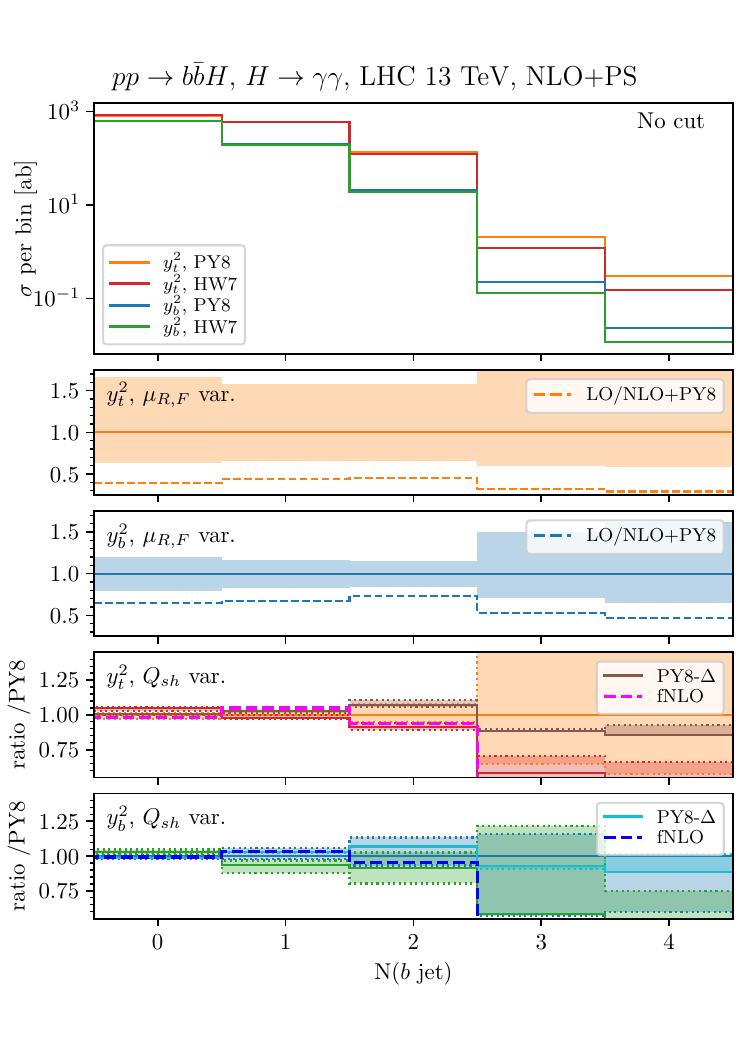}
    \includegraphics[width=0.48\textwidth]{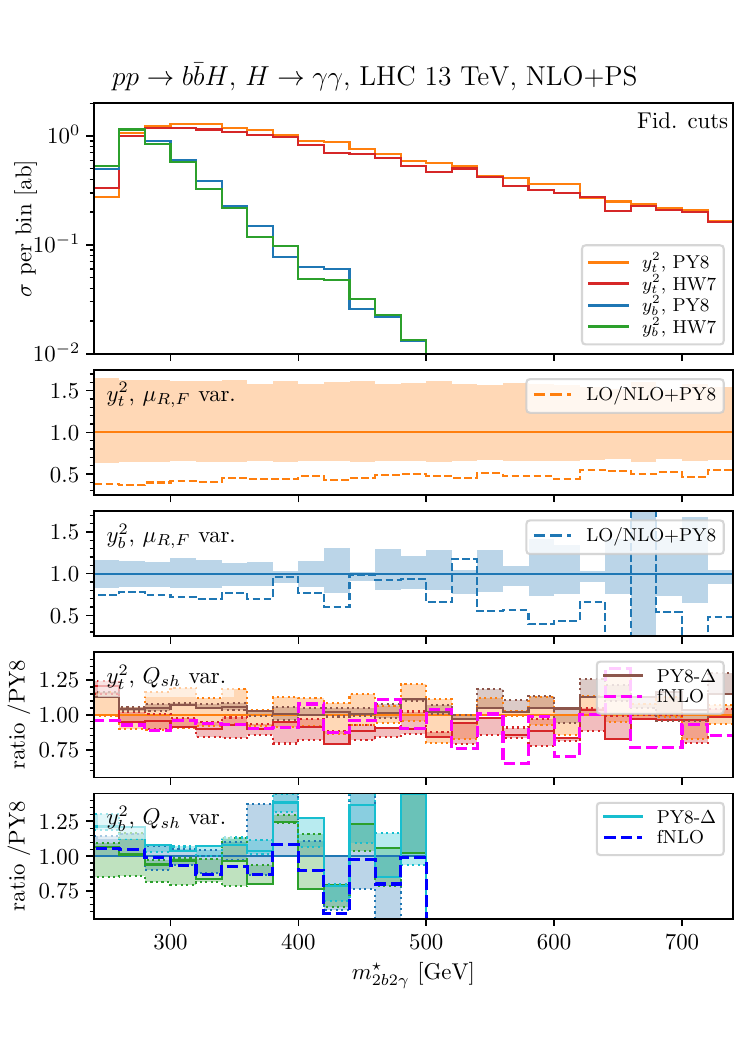}
    \vspace{-0.3cm}
    \caption{\label{fig:Nbm2b2g}Distribution in the number of $b$-jets without further cuts (left)
and in the $\mbbggs{}$ variable defined in \eqn{eq:mbbggs} with fiducial baseline
selection cuts applied (right).}
\end{figure}

We start by discussing the exclusive cross section as a function of the 
number of $b$-jets ($N_{b\textrm{-jet}}$) in the fully inclusive phase space in \fig{fig:Nbm2b2g}\,(left). 
From the main frame we can see that, while being similarly large for $N_{b\textrm{-jet}}=0$,
the \ytsq{} cross section becomes increasingly larger compared to the \ybsq{} one
with $N_{b\textrm{-jet}}$. This has already been found in \citere{Deutschmann:2018avk}, 
which concluded that requiring less (or softer) $b$-jets would increase the 
relative size of the \ybsq{} contribution.
As observed for the total inclusive cross section discussed in \tab{tab:rates}, the 
\ytsq{} contribution features even larger NLO QCD corrections than the \ybsq{} one.
This can be observed from the first two ratio panels by comparing the NLO+PS predictions
with the LO+PS curves. As a result, the \ytsq{} contribution is subject to a
much larger uncertainty stemming from variations of the perturbative scales.
This underlines again the crucial importance of NLO QCD corrections for \bbH{} production,
especially in the case of the component originating from the gluon-fusion process
proportional to \ytsq{}. Looking at the last two ratio panels, it is clear that up to 
$N_{b\textrm{-jet}}=2$, i.e.\ the multiplicities described at NLO+PS accuracy, 
shower-scale uncertainties are subleading. Only at higher multiplicities, which are 
effectively described only by the parton shower for $N_{b\textrm{-jet}}> 3$, the dependence on the shower scale becomes
significantly larger, as expected. Comparing the different NLO-matched predictions in these two ratio panels, we observe differences that increase with the 
$b$-jet multiplicity, both for the \ybsq{} and the \ytsq{} contributions, with {\sc Pythia8} in conjunction with MC@NLO-$\Delta$ displaying the hardest spectra,
while {\sc Herwig7} shows the softest ones, with the latter also being closer to the fixed-order predictions. However, differences 
are rather mild: if we consider the bin with $N_{b\textrm{-jet}}=2$, which is the one relevant for the fiducial cuts employed, they amount to about 20\% for both the \ytsq{} and the \ybsq{} contribution (being slightly larger for the latter) when the most different predictions are considered. Indeed,
the results in the fiducial phase space in \tab{tab:rates} display a very similar pattern.
Similar results have been observed for other processes featuring bottom quarks and heavier objects~\cite{Wiesemann:2014ioa,Degrande:2015vpa,Bagnaschi:2018dnh}. 
Considering the relatively large perturbative scale variations, the predictions 
from {\sc Pythia8} (standard and MC@NLO-$\Delta$) and {\sc Herwig7} are compatible within their respective uncertainties,
at least in the NLO+PS accurate multiplicities $N_{b\textrm{-jet}}\le 2$.

\begin{figure}[t]
    \centering
    \includegraphics[width=0.48\textwidth]{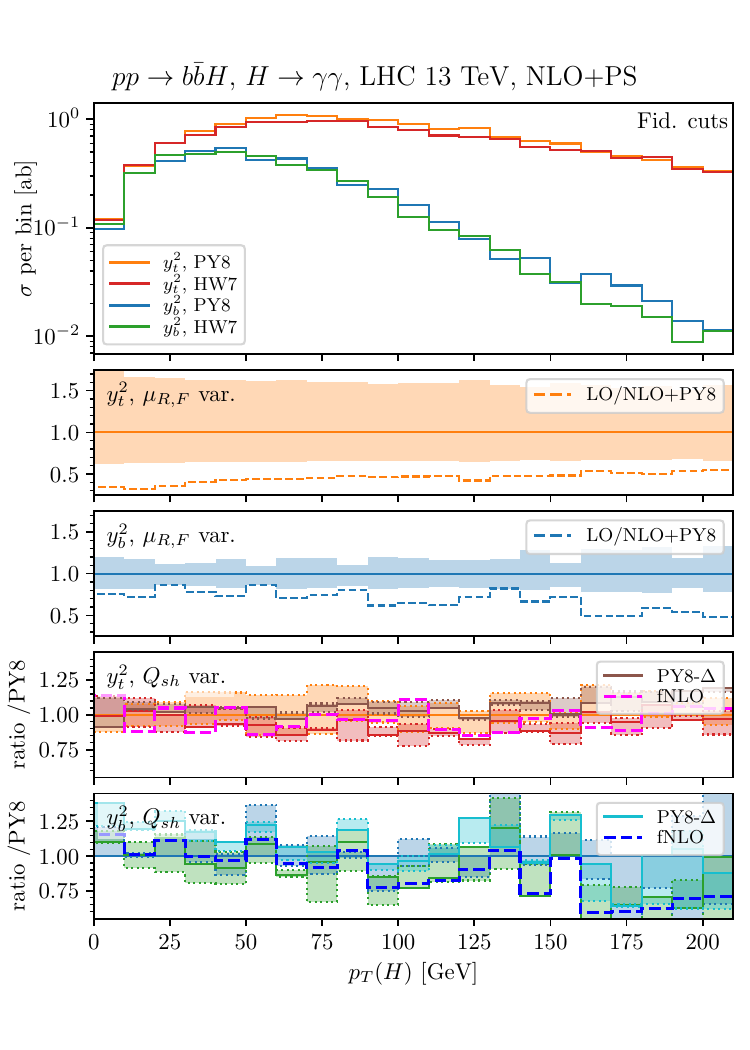}
     \includegraphics[width=0.48\textwidth]{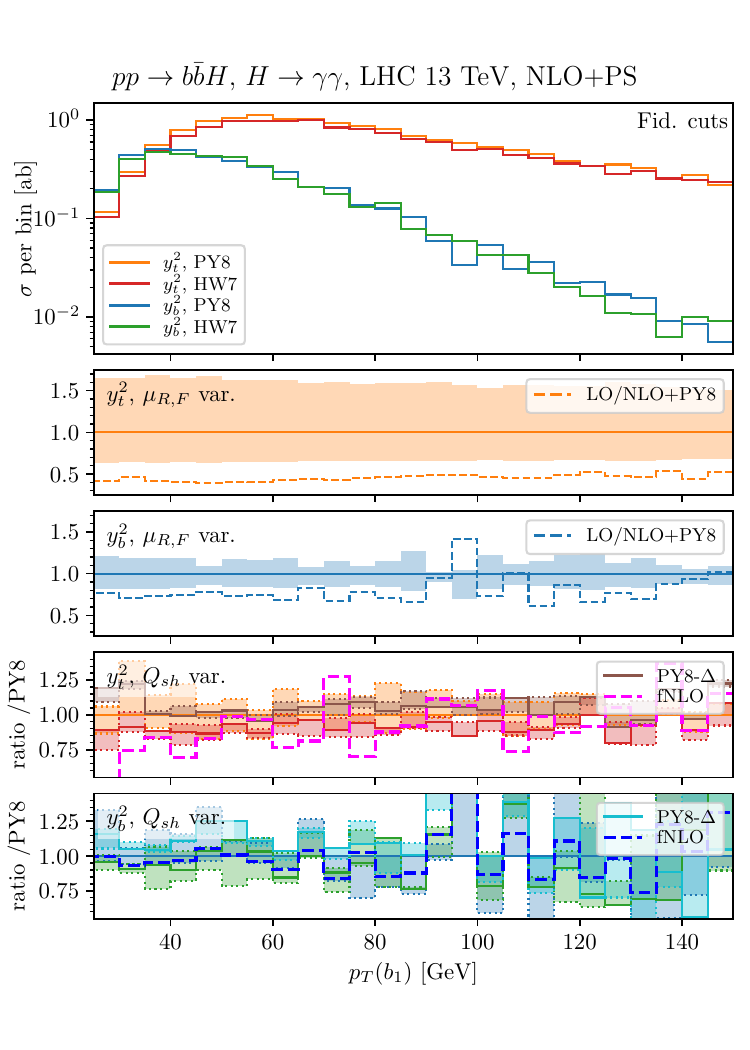}
     \vspace{-0.3cm}
     \caption{\label{fig:pthptb}Distribution in the transverse momentum of the Higgs boson reconstructed from the two photons (left) and of the leading $b$-jet (right), with
the fiducial baseline selection cuts applied.}
\end{figure}

Next, we consider the $\mbbggs{}$ distribution in \fig{fig:Nbm2b2g}\,(right) in the fiducial phase space.\footnote{Because of the much steeper decrease of the differential cross
section in the $\ybsq$ contribution, we show it only in a subset of the plotting range, up to $\mbbggs=500$ GeV.}
Also here we can appreciate from the main frame that the \ybsq{} contribution is 
strongly suppressed with increasing $\mbbggs{}$, while it becomes 
similarly large as the \ytsq{} one for $\mbbggs{}\lesssim 300$\,GeV.
This behaviour is also confirmed by the rates quoted in \tab{tab:rates}, where 
the relative size of the \ybsq{} contribution is largest in the last category 
with $\mbbggs{}< 350$\,GeV. This behaviour (reappearing in various kinematical 
distributions) originates from different features of the \ybsq{} contribution: firstly, the two bottom quarks are in general less correlated; secondly, the bottom quarks are predominantly
generated from initial-state splittings; thirdly, radiation off bottom quarks tends to be soft. 
Instead, for the \ytsq{} contribution the bottom quarks are dominantly produced by a hard gluon which recoils against the Higgs boson. Indeed, we can observe the suppression of the \ybsq{} contribution also 
at large transverse momenta of the Higgs boson $\pth$ and of the hardest $b$-jet 
$\ptbone$, shown in \fig{fig:pthptb}. 
As far as the scale variations and NLO QCD corrections visible in the first two 
ratio panels is concerned, we find very similar results as before for the number of 
$b$-jet also for $\mbbggs{}$, $\pth$ and $\ptbone$: the size of both 
the uncertainties and of the NLO corrections is larger for the \ytsq{} contribution.
We also notice that the relative behaviour of the various parton-shower predictions and the fixed order prediction
is in general not the same for the \ybsq{} and \ytsq{} contributions.

\subsection{Comparison and combination with the NNLOPS prediction}
\label{sec:comparison-with-NNLOPS}
	
\begin{figure}
    \centering
    \includegraphics[width=0.7\textwidth]{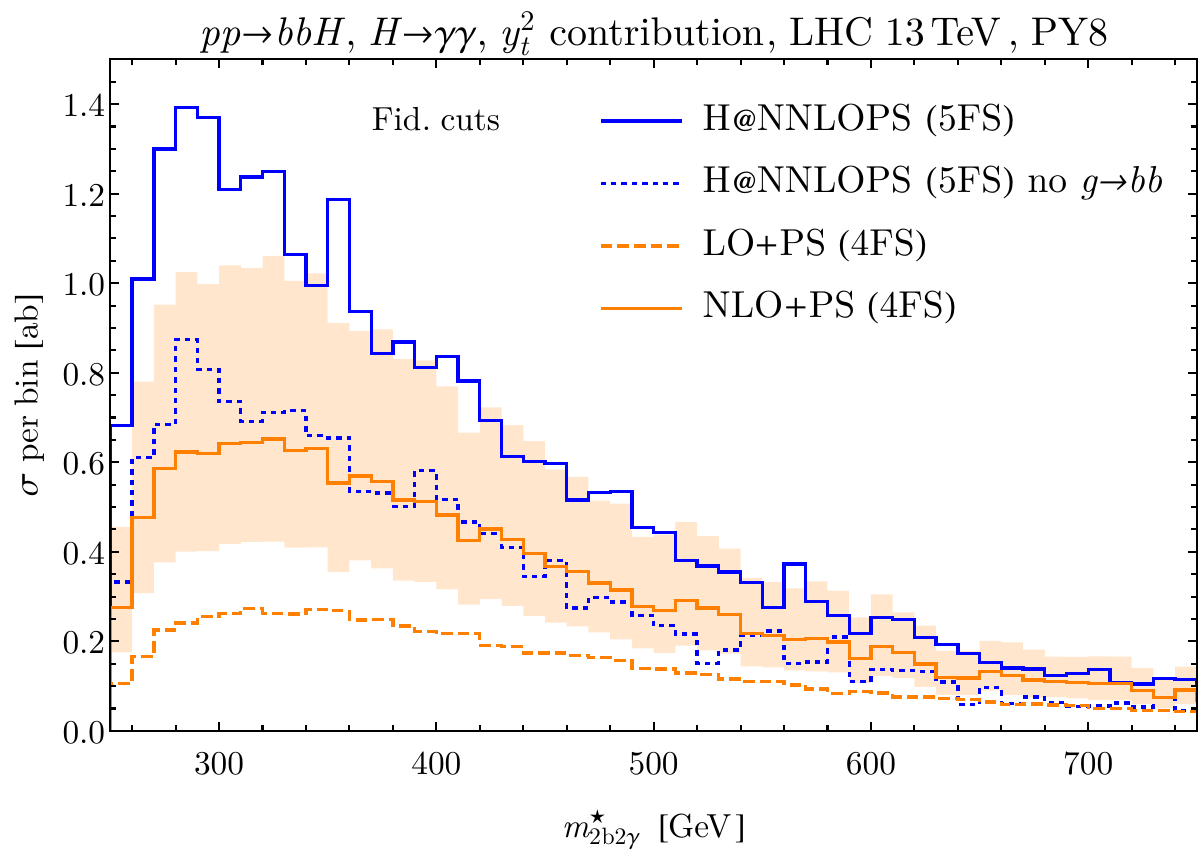}
    \vspace*{0.3cm}
\caption{\label{fig:compNNLOPS}Distribution in $m^\star_{2b2\gamma}$ with fiducial selection cuts, for the $y_t^2$ contribution. Predictions at LO and NLO in the 4FS are shown, as well as 5FS results, the latter with and without contributions from $g\to b\bar{b}$ splittings in the shower.}
\end{figure}

We now come back to the comparison of our 4FS NLO+PS predictions with results
from the 5FS NNLOPS generator used so far to model the \ytsq{}-induced
$b\bar{b}H$ background in $HH$ measurements. In \fig{fig:compNNLOPS} we consider the $\mbbggs$ distribution, which is used to define different fiducial categories by 
 the experiments, and show the
NNLOPS prediction with (blue solid curve) and without $g\to b\bar{b}$ splittings in the PS (blue dotted curve) 
together with our LO+PS (orange dashed curve) and NLO+PS (orange solid curve with orange band) predictions
in the 4FS. We immediately notice that the NNLOPS prediction is substantially larger than 
the 4FS prediction, especially at low $\mbbggs$ where it is even outside the orange uncertainty band. Moreover,
 the NNLOPS prediction reduces drastically and becomes more compatible with the NLO+PS
4FS prediction when secondary $g\to b\bar{b}$ splittings generated by the PS are turned off.
This is in line without our findings for the fiducial cross section in \tab{tab:rates}. 
Still, it is somewhat surprising that the LO-accurate 5FS result is as large as the NLO-accurate 
4FS prediction (at low $\mbbggs$ even larger), when the NLO corrections are of the order
of 100\%. 
One should bear in mind, however, that the scale setting of the NNLOPS prediction
is quite different. Since it is the typical scale for the inclusive production of a Higgs boson 
(not exclusive in the two bottom quarks), $\mu_R=\mu_F=m_H/2$ is used for the NNLO
prediction.
We also observe
that the 5FS predictions tend more towards small $\mbbggs$, which is not unexpected due to the 
fact that the bottom quarks are massless in the matrix elements and only put on the mass shell
through reshuffling of their momenta in the PS matching, which induces some arbitrariness
in their kinematics.

To further investigate why $g\to b\bar{b}$ splittings in the PS lead to such a substantial
and unexpected increase of the cross section with two hard $b$-jets, we will try to understand 
the type of topologies/kinematics that lead to this large contribution. First of all,
the PS acts somewhat outside its validity range if it creates two hard $b$-jets, unless these
originate from two hard jets each splitting into a collinear $b$-jet pair, thus giving rise to four bottom quarks in the event. We have checked 
that in the relevant phase space region the vast majority of events have exactly 
two bottom quarks in total. Therefore, the PS creates two $b$-jets at large angle out of a 
single $g\to b\bar{b}$ splitting, which in principle should barely happen and is very poorly 
described in the soft/collinear approximation of the PS.
In this situation, there are only two possible 
configurations that could lead to two hard $b$-jets. In the first one, two soft and wide-angle bottom quarks 
are created, which the jet-clustering algorithm combines with other hard partons to 
form $b$-jets. In other words, in these configurations, only a small fraction of the jet energy is carried by the bottom quarks. Such an effect could be missing in our 4FS $b\bar{b}H$ calculation
at NLO, as we generally include two hard $b$-jets and (at most) one additional hard light jet
at matrix-element level. All other jets are generated by the PS in our calculation 
and should be softer. The relevant hard matrix element for this configuration has two bottom quarks
and two light jets in the final state ($pp\to b\bar{b}Hjj$), which enters beyond the accuracy of our calculation, but, in principle, could be included through multi-jet merging.
In the second possible configuration, the PS creates two hard wide-angle bottom quarks, each of which creates a separate
$b$-jet subsequently. In this case, the bottom quark will carry a rather large fraction of the $b$-jet energy. Should this configuration be the dominant one, then not only would the PS describe such kinematics very poorly, but also the 
relevant hard matrix-element for such configurations is simply $gg\to b\bar{b}H$ (and possibly 
further soft partons). This contribution is actually already accounted for in the LO matrix element
of the 5FS NNLOPS calculation and should not be created again by the PS.

\begin{figure}
    \centering
    \includegraphics[width=0.48\textwidth]{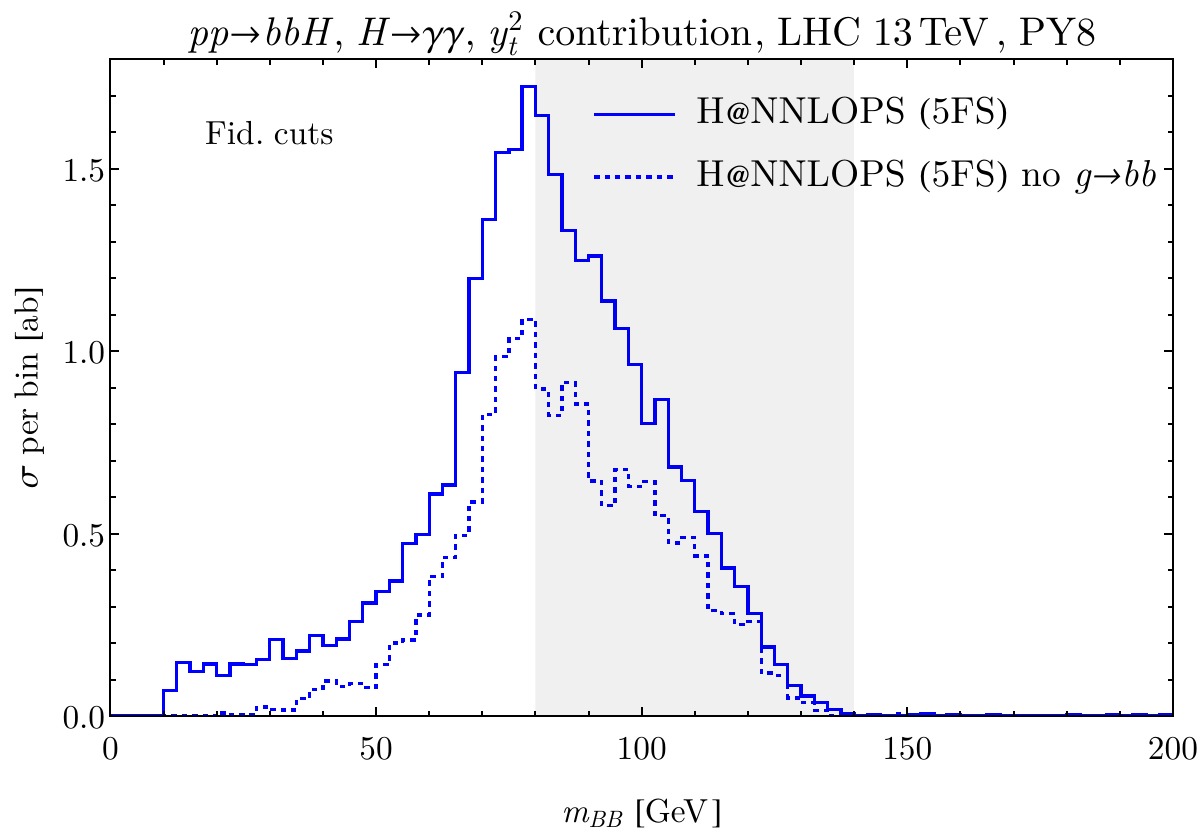}
 \includegraphics[width=0.48\textwidth]{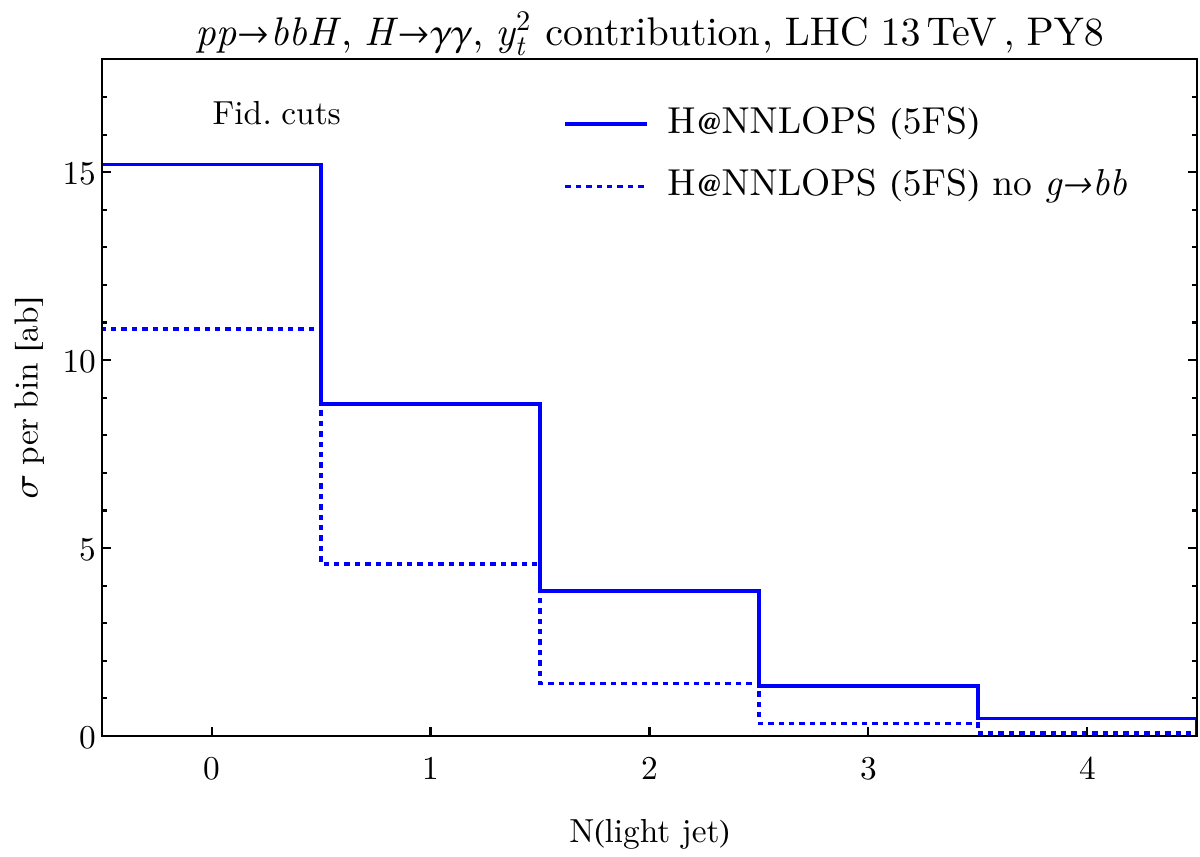}
 \vspace*{0.3cm}
\caption{\label{fig:compNNLOPS2}Distribution in the invariant mass of the two hardest $B$ hadrons (left) and in the number of light jets (right) in the 5FS calculation, with and without the contributions from $g\to b\bar{b}$ splittings in the shower, in the fiducial region. The shaded area in the left-hand side plot corresponds to the invariant-mass requirement on the $b$-jets employed in our fiducial cuts.}
\end{figure}

In order to asses which of the two aforementioned configurations is responsible for the observed large effect due to  
$g\to b\bar{b}$ splittings in the PS, we show two further distributions in 
\fig{fig:compNNLOPS2}, namely the invariant mass of the two hardest $B$ hadrons in the signal 
region and the number of light jets, in both cases turning on and off $g\to b\bar{b}$ splittings 
in the PS. What we can see is that the PS creates two hard $B$ hadrons 
at wide angle that form two separate $b$-jets. Their invariant mass is quite close to the window required for the two $b$-jets in \eqn{eq:mbjcut}, displayed in the figure as a shaded region. Thus, we can rule out
the first scenario. From the distribution in the number of light jets
we can see that more than half of the events have no extra light jets,
and that in the contribution coming exclusively from $g\to b\bar{b}$
splittings (obtained as the difference between the two curves in \fig{fig:compNNLOPS2})
events with no extra light jets represent a large fraction (more than one third) of the total.
All this strongly points to the fact that the contribution
due to secondary $g\to b\bar{b}$ splittings 
is counted twice (since it is already accounted for by the $gg\to b\bar{b}H$ hard matrix 
element), and that the PS is acting beyond its validity range and generates two hard and separated
bottom quarks. Moreover, events with two or more light jets represent less than one third of the contribution
from $g\to b\bar{b}$ splittings. Such a contribution may actually be missing in our 4FS calculation, 
however, it should be well covered by the $\sim +60\%$ higher-order 
uncertainties on the predictions.

Before concluding, we would like to point out that the NNLOPS calculation is used
in experimental analyses not only as an estimate of the \bbH{} \ytsq{} background, as discussed
at length in this work, but also to model the contribution
originating from inclusive Higgs production with fake $b$-jets, i.e.\ where light jets
are mistagged as $b$-jets.
Therefore, even if our NLO+PS description of the \bbH{} background in the 4FS is to be
  used, replacing the NNLOPS calculation for the \ytsq{} contribution in this case, a consistent way
  of combining both simulations needs to be devised to be
able to simulate the fake component as well.
The most simple and naive option, which already provides a largely consistent
prediction, can be achieved at the level of the events, without the need of regenerating the event samples:
In the 5FS NNLOPS events, all contributions with final-state bottom quarks (irrespective
whether they originate from the hard matrix element, from initial-state splittings of 
bottom quarks, or final-state $g\to b\bar{b}$ splittings generated by the shower) would have to be 
removed from the final result. This would cancel the \bbH{} background entirely from the 
5FS NNLOPS sample, and only the contributions from light jets (and potential fakes)
would be kept. On top of this, the contribution from 4FS NLO+PS \bbH{} events can be 
added incoherently. The same approach has been used in \citere{Bagnaschi:2018dnh} to estimate 
bottom mass effects in the $Z$-boson transverse momemtum.
The only drawback of this approach is that formally a 5FS and a 4FS calculation are mixed. 
Thus, to be fully consistent, one would have to rerun the NNLOPS calculation with 4FS
PDF sets and strong coupling, as well as setting the number of light flavours to $n_f=4$ 
in the calculation. 
In this case, events with initial- or final-state bottom quarks (at the level of the hard matrix element) would
be removed from the calculation by construction. This can be achieved by removing 
the appropriate flavour structures from the original 5FS implementation 
of the NNLOPS generator.
Moreover, the $g\to b\bar{b}$ splittings will have to be turned off when showering the NNLOPS
events, to avoid double counting with the 4FS NLO+PS \bbH{} calculation.
As a result, one obtains  a consistent 4FS prediction including both NNLO QCD accuracy 
to inclusive Higgs production and NLO QCD accuracy for the \bbH{} background,
including their matching to the parton shower.

\subsection[Impact of the new $\rm b\bar{b}H$ modeling for the $\rm HH$ searches]{Impact of the new \boldmath{$\rm b\bar{b}H$} modeling for the \boldmath{$\rm HH$} searches}
\label{sec:impact-on-hh-searches}	

We have estimated the impact of the NLO QCD modeling of the \ytsq{} contribution with respect to previously adopted NNLOPS prediction on the current limits for the $HH$ cross section (and $\lambda_{HHH}$). 
The rates and the uncertainties from the new \bbH{} prediction are propagated to the search for $HH$ production in the $2b2\gamma$ final state performed by the ATLAS experiment by using the publicly available information in \citere{ATLAS:2021ifb}.

The inclusive $gg\to H$ background involves both the $gg\to \bbH{}$ process and Higgs boson production with additional jets where at least one is mistagged as a $b$-jet (i.e.\ fake $b$-jets). The decomposition of the $gg\to H$ background in terms of the flavour of the two additional jets is estimated by evaluating the rates of the $gg\to H$ plus $b$-jets, $c$-jets, or light jets contributions from the NNLOPS prediction, and multiplying each with the appropriate $b$-tagging (or mistagging) efficiency of the $b$-tagging algorithm employed in the $HH\to2b2\gamma$ search~\cite{ATLAS:2022qxm}.
This results in 80\% (20\%) of the inclusive $gg\to H$ background arising from events with true (fake) $b$-jets.
To assess the impact of the new \bbH{} calculation, we rescaled only the 80\% of the inclusive $gg\to H$ background estimated by the analysis with the rates from the new \ytsq-induced \bbH{} calculation quoted in Table~\ref{tab:rates}.
Furthermore, the 100\% uncertainty assigned to the $gg\to H$ process in the current analysis is replaced with the uncertainty arising from the scale variations of the \ytsq-induced $\bbH{}$ component reported in Table~\ref{tab:rates}.

Following the procedure described above, we have estimated that new \bbH{} modeling improves the limit on the $HH$ production cross section by a few percent. The relatively small improvement is mostly due to the fact that the $gg\to H$ process is not the main contribution to the background for the $HH\to 2b2\gamma$ search, which is dominated by the SM production of $\gamma\gamma jj$ final states. Moreover, the current analysis, based on Run 2 data, corresponding to a total integrated luminosity of 140 fb$^{-1}$, is largely limited by statistical uncertainties.
Similarly, we have studied the effect of the
new \bbH{} modeling on $HH$ searches during the 
High-Luminosity phase of the LHC by projecting these results to 3~ab$^{-1}$ integrated luminosity, as done in \citere{ATL-PHYS-PUB-2022-001}. 
At the HL-LHC, the 100\% systematic uncertainty on the $gg\rightarrow H$ background starts to be one of the limiting factors for the sensitivity of the analysis. In this case, the improvement coming from the new \bbH{} modeling increases to around 5\% for the limit on both the $HH$ production cross section and $\lambda_{HHH}$, as well as for the $HH$ discovery significance.

Finally, we have applied the same recipe to estimate the effect of the new $\bbH{}$ modeling on the $HH$ search in the $2b2\tau$ channel~\cite{ATLAS:2022xzm}. The same flavour decomposition found for the $2b2\gamma$ final state is assumed for the $2b2\tau$ channel, and so is the contribution of the $gg\to H$ process to the inclusive single Higgs background with respect to the other single Higgs production modes. In addition, the fiducial region defined by the $2b2\tau$ analysis was not studied in detail. Instead, the same \bbH{} rates showed in Table~\ref{tab:rates}, have been assumed.
For the $2b2\tau$ final state, employing the new \bbH{} prediction improves the limit on the $HH$ production cross section by up to 10\%, and, after extrapolating the analysis to the HL-LHC scenario, the improvement on both the discovery significance and the $HH$ cross section limit increases up to $20\%$.
The larger effect seen on the $2b2\tau$ final state is mainly due to the fact that the analysis is less dominated by the statistical uncertainty and that the single Higgs background is more relevant with respect to the $2b2\gamma$ channel.

\section{Conclusions}
\label{sec:conclusions}
We have presented a new study of \bbH{} production in the SM, focusing
on its impact in the Higgs-pair production signal region. 
The dominant contributions to the \bbH{} process,
namely those proportional to \ybsq{} and \ytsq{}, are simulated at NLO in the 4FS and matched to parton showers by employing the {\sc MadGraph5\_aMC@NLO} framework. This is the first calculation of this kind for the latter contribution. 
The accurate modeling of the \bbH{} background to Higgs-pair measurements is particularly relevant, since in all sensitive search channels at least 
one of the two Higgs bosons decays to bottom quarks, rendering the continuum 
production of the \bbH{} final state an irreducible and very sizeable background. In fact,
due to the low accuracy in the simulation of the \ytsq{} contribution used in the current 
experimental searches, \bbH{} production yields the dominant theoretical 
uncertainty in $HH$ measurements.

In this paper, we have studied in detail the impact of including the full 
NLO QCD corrections to the \bbH{} process in the context of $HH$ production.
We found that corrections of $\mathcal{O}(+100\%)$ and larger increase the
\ytsq{} contribution at NLO in QCD, which renders it vital to include these effects
for a sufficiently accurate modeling of the \bbH{} background in $HH$ analyses.
Although the total uncertainties (from factorization-scale, renormalization-scale, shower-scale and shower variations) are still relatively large (at the level of +50\%
and -30\%), this constitutes a substantial improvement over the 
previously employed LO approximation for the \ytsq{} contribution to the \bbH{} background 
from the NNLOPS generator 
with $\mathcal{O}(100\%)$ uncertainties, not only 
in terms of accuracy, but also in terms of precision.

When comparing our results to the previously employed 5FS NNLOPS calculation,
which describes the \ytsq{} \bbH{} background effectively with LO+PS accuracy, we notice 
that the NNLOPS prediction is almost a factor of two larger than our NLO+PS
prediction, despite missing the $\mathcal{O}(+100\%)$  NLO correction.
We have traced this back to two major sources: Firstly, contributions with 
two hard bottom quarks generated from a single wide-angle $g\to b\bar{b}$ splitting in
the parton shower yield about half of the 5FS NNLOPS prediction.
Not only does the parton shower act vastly beyond its validity range in this case,
but also a large part of these configurations have no extra light jets and 
 should already be accounted for by the hard LO
matrix element, which suggests that there is some kind of double counting in this
 kinematical regime. Secondly, the scale setting in the NNLOPS sample adopted 
 for inclusive Higgs production, which is of the order of the Higgs mass, 
 does not really reflect the hardness of the \bbH{} final state.
 With this different scale setting the LO+PS 5FS result (without $g\to b\bar{b}$ splittings from the shower) is as large as our 4FS NLO+PS prediction.
We note that the issues with the 5FS description may affect also other processes including bottom quarks in the final state.
 
We have estimated the potential impact of accurate predictions for the \bbH{} background for $HH$ searches. 
We concluded that the current limits on $\sigma^{HH}_{\rm SM}$ could improve by few percent and up to 10\% in the $2b2\gamma$ and $2b2\tau$ channels, respectively. 
At HL-LHC the improvements on the $\sigma^{HH}_{\rm SM}$ limits and on the $HH$ discovery significance increases to 5\% (20\%) for the $2b2\gamma$ ($2b2\tau$) final state.
Given that the current limit is several times the SM value of the $HH$ cross section,
this is not a particularly dramatic change. However, since with the end of the 
High-Luminosity phase of the LHC Higgs pair production is expected to be observed
at almost five standard deviations, the accurate modeling of the \bbH{} background, 
which we found to be as large as the $HH$ cross section in the relevant fiducial 
phase space regions, will be indispensable for maximizing the sensitivity to the 
$HH$ signal. Moreover, it will ensure that bounds/extraction of the $HH$ cross section
as well as the trilinear Higgs coupling will be fully reliable. In addition, an improved description of the \bbH{} process can be beneficial not only in the context of $HH$ physics, but also in analyses targeting single-Higgs production, such as measurements of Higgs boson production in association with a top quark pair, where \bbH{} production also contributes as a background~\cite{ATLAS:2018mme}.
For all of these reasons, the \bbH{} generator presented in this work will represent a very useful tool for future analyses performed by the LHC collaborations.
In principle, also the inclusion of NNLO
QCD corrections to this type of process is feasible, which is left for future work.

\noindent {\bf Acknowledgements.}
We would like to thank Paolo Nason for clarifications about the NNLOPS generator and
Roy Stegeman for providing us with the previous version of the NNPDF set employed in
this paper for cross checks.
MW thanks Rhorry Gauld and Giulia Zanderighi for fruitful 
discussions on the topic.
EM is indebted to Leonardo Carminati and Ruggero Turra for many discussions and
constant support.
MZ~is supported by the ``Programma per Giovani Ricercatori Rita Levi Montalcini'' granted by the Italian Ministero dell'Universit\`a e della Ricerca (MUR). 

\appendix

\setlength{\bibsep}{3.1pt}
\renewcommand{\em}{}
\normalem
\bibliographystyle{JHEP}
\bibliography{bbH_as_HH_background}
\end{document}

%% file: definitions.tex
\providecommand{\href}[2]{#2}

\newcommand\as{\alpha_{\mathrm{S}}}

\def\to{\rightarrow}

\def\mbbggs{m_{2b2\gamma}^{\star}}

\def\perc{\%}

\def\GeV{\mathrm{GeV}}

\newcommand{\eqn}[1]{Eq.\,(\ref{#1})}

\newcommand{\fig}[1]{Figure~\ref{#1}}

\newcommand{\tab}[1]{Table~\ref{#1}}
\newcommand{\sct}[1]{Section~\ref{#1}}

\def\citere#1{\mbox{Ref.~\cite{#1}}}
\def\citeres#1{\mbox{Refs.~\cite{#1}}}

\usepackage{etoolbox}
\makeatletter
\patchcmd{\@sect}{#8}{\boldmath #8}{}{}
\let\ori@chapter\@chapter
\def\@chapter[#1]#2{\ori@chapter[\boldmath#1]{\boldmath#2}}
\makeatother

\newcommand\calo[1]{{\cal O}\hspace{-0.2em}\left(#1\right)}

\newcommand{\bbH}{\ensuremath{b\bar{b}H}}

\newcommand{\yt}{\ensuremath{y_t}}
\newcommand{\ytsq}{\ensuremath{y_t^2}}
\newcommand{\yb}{\ensuremath{y_b}}
\newcommand{\ybsq}{\ensuremath{y_b^2}}
\newcommand{\ybyt}{\ensuremath{y_b\, y_t}}

\newcommand{\pth}{\ensuremath{p_T(H)}}

\newcommand{\ptbone}{\ensuremath{p_T(b_1)}}

%% file: table_rates_inc.tex
$y_b^2$  &   247  & $^{+54\perc} _{-33\perc}$  &   374  & $^{+18\perc} _{-20\perc}$\\
$y_t^2$  &   289  & $^{+69\perc} _{-38\perc}$  &   689  & $^{+61\perc} _{-35\perc}$\\
sum  &   536  & $^{+62\perc} _{-35\perc}$  &  1064  & $^{+46\perc} _{-29\perc}$\\

%% file: table_rates.tex
\multirow{9}{*}{\makecell{No cut}}  
& \multirow{3}{*}{$y_b^2$}  
&  \cellcolor{White}                      PY8  &  \cellcolor{White} ${ 561}$  & \cellcolor{White} ${ 849}$  & \cellcolor{White} $^{+18\perc} _{-20\perc}$  & \cellcolor{White} $^{+0\perc} _{+0\perc}$   &  \multirow{9}{*}{\makecell{4867 \\ \\ $\cancel{g\to b\bar{b}}$:\\ 2140}}  &  \multirow{9}{*}{$ 82.1 $} \\
&  &  \cellcolor{Gray!15}             PY8-$\Delta$  &   \cellcolor{Gray!15} ${}$  & \cellcolor{Gray!15} ${ 848}$  & \cellcolor{Gray!15} & \cellcolor{Gray!15} $^{+0\perc} _{+0\perc}$    & & \\
&  &  \cellcolor{Gray!30}                      HW7  &   \cellcolor{Gray!30} ${ 561}$  & \cellcolor{Gray!30} ${ 851}$  & \cellcolor{Gray!30} & \cellcolor{Gray!30} $^{+0\perc} _{+0\perc}$    & & \\
& \multirow{3}{*}{$y_t^2$}  
&  \cellcolor{White}                      PY8  &  \cellcolor{White} ${ 655}$  & \cellcolor{White} ${1565}$  & \cellcolor{White} $^{+61\perc} _{-35\perc}$  & \cellcolor{White} $^{+0\perc} _{+0\perc}$    & & \\
&  &  \cellcolor{Gray!15}             PY8-$\Delta$  &   \cellcolor{Gray!15} ${}$  & \cellcolor{Gray!15} ${1595}$  & \cellcolor{Gray!15} & \cellcolor{Gray!15} $^{+0\perc} _{+0\perc}$    & & \\
&  &  \cellcolor{Gray!30}                      HW7  &   \cellcolor{Gray!30} ${ 655}$  & \cellcolor{Gray!30} ${1578}$  & \cellcolor{Gray!30} & \cellcolor{Gray!30} $^{+0\perc} _{+0\perc}$    & & \\
& \multirow{3}{*}{sum}  
&  \cellcolor{White}                      PY8  &  \cellcolor{White} ${1217}$  & \cellcolor{White} ${2414}$  & \cellcolor{White} $^{+46\perc} _{-29\perc}$  & \cellcolor{White} $^{+0\perc} _{+0\perc}$    & & \\
&  &  \cellcolor{Gray!15}             PY8-$\Delta$  &   \cellcolor{Gray!15} ${}$  & \cellcolor{Gray!15} ${2443}$  & \cellcolor{Gray!15} & \cellcolor{Gray!15} $^{+0\perc} _{+0\perc}$    & & \\
&  &  \cellcolor{Gray!30}                      HW7  &   \cellcolor{Gray!30} ${1216}$  & \cellcolor{Gray!30} ${2429}$  & \cellcolor{Gray!30} & \cellcolor{Gray!30} $^{+0\perc} _{+0\perc}$    & & \\
\hline
\multirow{9}{*}{\makecell{Fid. cuts}}  
& \multirow{3}{*}{$y_b^2$}  
&  \cellcolor{White}                      PY8  &  \cellcolor{White} ${3.15}$  & \cellcolor{White} ${4.22}$  & \cellcolor{White} $^{+15\perc} _{-15\perc}$  & \cellcolor{White} $^{+10\perc} _{-4\perc}$   &  \multirow{9}{*}{\makecell{29.9 \\ \\ $\cancel{g\to b\bar{b}}$:\\ 17.2}}  &  \multirow{9}{*}{$ 22.7 $} \\
&  &  \cellcolor{Gray!15}             PY8-$\Delta$  &   \cellcolor{Gray!15} ${}$  & \cellcolor{Gray!15} ${4.75}$  & \cellcolor{Gray!15} & \cellcolor{Gray!15} $^{+0\perc} _{-2\perc}$    & & \\
&  &  \cellcolor{Gray!30}                      HW7  &   \cellcolor{Gray!30} ${2.59}$  & \cellcolor{Gray!30} ${4.08}$  & \cellcolor{Gray!30} & \cellcolor{Gray!30} $^{+8\perc} _{-12\perc}$    & & \\
& \multirow{3}{*}{$y_t^2$}  
&  \cellcolor{White}                      PY8  &  \cellcolor{White} ${8.24}$  & \cellcolor{White} ${18.1}$  & \cellcolor{White} $^{+58\perc} _{-34\perc}$  & \cellcolor{White} $^{+10\perc} _{-7\perc}$    & & \\
&  &  \cellcolor{Gray!15}             PY8-$\Delta$  &   \cellcolor{Gray!15} ${}$  & \cellcolor{Gray!15} ${19.2}$  & \cellcolor{Gray!15} & \cellcolor{Gray!15} $^{+3\perc} _{-1\perc}$    & & \\
&  &  \cellcolor{Gray!30}                      HW7  &   \cellcolor{Gray!30} ${6.83}$  & \cellcolor{Gray!30} ${16.6}$  & \cellcolor{Gray!30} & \cellcolor{Gray!30} $^{+4\perc} _{-5\perc}$    & & \\
& \multirow{3}{*}{sum}  
&  \cellcolor{White}                      PY8  &  \cellcolor{White} ${11.4}$  & \cellcolor{White} ${22.3}$  & \cellcolor{White} $^{+50\perc} _{-30\perc}$  & \cellcolor{White} $^{+10\perc} _{-6\perc}$    & & \\
&  &  \cellcolor{Gray!15}             PY8-$\Delta$  &   \cellcolor{Gray!15} ${}$  & \cellcolor{Gray!15} ${23.9}$  & \cellcolor{Gray!15} & \cellcolor{Gray!15} $^{+2\perc} _{-1\perc}$    & & \\
&  &  \cellcolor{Gray!30}                      HW7  &   \cellcolor{Gray!30} ${9.42}$  & \cellcolor{Gray!30} ${20.7}$  & \cellcolor{Gray!30} & \cellcolor{Gray!30} $^{+4\perc} _{-6\perc}$    & & \\
\hline
\multirow{9}{*}{\makecell{Fid. cuts \\+ $\mbbggs<500\,\GeV$}}  
& \multirow{3}{*}{$y_b^2$}  
&  \cellcolor{White}                      PY8  &  \cellcolor{White} ${3.11}$  & \cellcolor{White} ${4.15}$  & \cellcolor{White} $^{+15\perc} _{-15\perc}$  & \cellcolor{White} $^{+11\perc} _{-4\perc}$   &  \multirow{9}{*}{\makecell{22.3 \\ \\ $\cancel{g\to b\bar{b}}$:\\ 13.3}}  &  \multirow{9}{*}{$ 15.7 $} \\
&  &  \cellcolor{Gray!15}             PY8-$\Delta$  &   \cellcolor{Gray!15} ${}$  & \cellcolor{Gray!15} ${4.69}$  & \cellcolor{Gray!15} & \cellcolor{Gray!15} $^{+0\perc} _{-2\perc}$    & & \\
&  &  \cellcolor{Gray!30}                      HW7  &   \cellcolor{Gray!30} ${2.56}$  & \cellcolor{Gray!30} ${4.02}$  & \cellcolor{Gray!30} & \cellcolor{Gray!30} $^{+8\perc} _{-13\perc}$    & & \\
& \multirow{3}{*}{$y_t^2$}  
&  \cellcolor{White}                      PY8  &  \cellcolor{White} ${5.33}$  & \cellcolor{White} ${12.3}$  & \cellcolor{White} $^{+60\perc} _{-34\perc}$  & \cellcolor{White} $^{+12\perc} _{-8\perc}$    & & \\
&  &  \cellcolor{Gray!15}             PY8-$\Delta$  &   \cellcolor{Gray!15} ${}$  & \cellcolor{Gray!15} ${12.8}$  & \cellcolor{Gray!15} & \cellcolor{Gray!15} $^{+2\perc} _{-1\perc}$    & & \\
&  &  \cellcolor{Gray!30}                      HW7  &   \cellcolor{Gray!30} ${4.31}$  & \cellcolor{Gray!30} ${11.3}$  & \cellcolor{Gray!30} & \cellcolor{Gray!30} $^{+5\perc} _{-5\perc}$    & & \\
& \multirow{3}{*}{sum}  
&  \cellcolor{White}                      PY8  &  \cellcolor{White} ${8.44}$  & \cellcolor{White} ${16.5}$  & \cellcolor{White} $^{+49\perc} _{-29\perc}$  & \cellcolor{White} $^{+12\perc} _{-7\perc}$    & & \\
&  &  \cellcolor{Gray!15}             PY8-$\Delta$  &   \cellcolor{Gray!15} ${}$  & \cellcolor{Gray!15} ${17.5}$  & \cellcolor{Gray!15} & \cellcolor{Gray!15} $^{+1\perc} _{-1\perc}$    & & \\
&  &  \cellcolor{Gray!30}                      HW7  &   \cellcolor{Gray!30} ${6.86}$  & \cellcolor{Gray!30} ${15.3}$  & \cellcolor{Gray!30} & \cellcolor{Gray!30} $^{+6\perc} _{-7\perc}$    & & \\
\hline
\multirow{9}{*}{\makecell{Fid. cuts \\+ $\mbbggs<350\,\GeV$}}  
& \multirow{3}{*}{$y_b^2$}  
&  \cellcolor{White}                      PY8  &  \cellcolor{White} ${2.71}$  & \cellcolor{White} ${3.65}$  & \cellcolor{White} $^{+15\perc} _{-16\perc}$  & \cellcolor{White} $^{+9\perc} _{-4\perc}$   &  \multirow{9}{*}{\makecell{11.5 \\ \\ $\cancel{g\to b\bar{b}}$:\\ 6.82}}  &  \multirow{9}{*}{$ 2.84 $} \\
&  &  \cellcolor{Gray!15}             PY8-$\Delta$  &   \cellcolor{Gray!15} ${}$  & \cellcolor{Gray!15} ${4.11}$  & \cellcolor{Gray!15} & \cellcolor{Gray!15} $^{+0\perc} _{-2\perc}$    & & \\
&  &  \cellcolor{Gray!30}                      HW7  &   \cellcolor{Gray!30} ${2.22}$  & \cellcolor{Gray!30} ${3.54}$  & \cellcolor{Gray!30} & \cellcolor{Gray!30} $^{+8\perc} _{-15\perc}$    & & \\
& \multirow{3}{*}{$y_t^2$}  
&  \cellcolor{White}                      PY8  &  \cellcolor{White} ${2.32}$  & \cellcolor{White} ${5.78}$  & \cellcolor{White} $^{+61\perc} _{-34\perc}$  & \cellcolor{White} $^{+13\perc} _{-9\perc}$    & & \\
&  &  \cellcolor{Gray!15}             PY8-$\Delta$  &   \cellcolor{Gray!15} ${}$  & \cellcolor{Gray!15} ${6.05}$  & \cellcolor{Gray!15} & \cellcolor{Gray!15} $^{+1\perc} _{+0\perc}$    & & \\
&  &  \cellcolor{Gray!30}                      HW7  &   \cellcolor{Gray!30} ${1.88}$  & \cellcolor{Gray!30} ${5.43}$  & \cellcolor{Gray!30} & \cellcolor{Gray!30} $^{+5\perc} _{-3\perc}$    & & \\
& \multirow{3}{*}{sum}  
&  \cellcolor{White}                      PY8  &  \cellcolor{White} ${5.03}$  & \cellcolor{White} ${9.43}$  & \cellcolor{White} $^{+44\perc} _{-27\perc}$  & \cellcolor{White} $^{+12\perc} _{-7\perc}$    & & \\
&  &  \cellcolor{Gray!15}             PY8-$\Delta$  &   \cellcolor{Gray!15} ${}$  & \cellcolor{Gray!15} ${10.2}$  & \cellcolor{Gray!15} & \cellcolor{Gray!15} $^{+0\perc} _{+0\perc}$    & & \\
&  &  \cellcolor{Gray!30}                      HW7  &   \cellcolor{Gray!30} ${4.10}$  & \cellcolor{Gray!30} ${8.97}$  & \cellcolor{Gray!30} & \cellcolor{Gray!30} $^{+6\perc} _{-8\perc}$    & & \\
\hline